\documentclass[a4paper]{jpconf}
\usepackage{graphicx}

\usepackage{hyperref}
\usepackage{xspace}
\usepackage{lineno}

\begin{document}
\title{Origin of atmospheric aerosols at the Pierre Auger Observatory using backward trajectory of air masses}

\author{Karim Louedec$^{1}$, for the Pierre Auger Collaboration$^{2}$}

\address{$^1$Laboratoire de Physique Subatomique et de Cosmologie (LPSC), UJF--INPG, CNRS/IN2P3, Grenoble, France.}
\address{$^2$Observatorio Pierre Auger, Av. San Mart\'in Norte 304, 5613 Malarg\"ue, Argentina. \\(Full author list: http://www.auger.org/archive/authors\_2012\_04.html)}

\ead{karim.louedec@lpsc.in2p3.fr}

\begin{abstract}
The Pierre Auger Observatory is the largest operating cosmic ray observatory ever built. Calorimetric measurements of extensive air showers induced by cosmic rays are performed with a fluorescence detector. Thus, one of the main challenges is the monitoring of the atmosphere, both in terms of atmospheric state variables and optical properties. To better understand the atmospheric conditions, a study of air mass trajectories above the site is presented. Such a study has been done using an air-modelling program well known in atmospheric sciences. Its validity has been checked using meteorological radiosonde soundings performed at the Pierre Auger Observatory. Finally, aerosol concentration values measured by the Central Laser Facility are compared to backward trajectories.
\end{abstract}

The Pierre Auger Observatory~\cite{PAO_1} measures the flux, arrival direction distribution and mass composition of cosmic rays from $10^{18}~$eV to the very highest energies. It is located in the ``Pampa Amarilla'' ($35.1^\circ - 35.5^\circ~${\bf S}, $69.0^\circ - 69.6^\circ~${\bf W}, and $1300-1400~$m above sea level), close to Malarg\"{u}e, Province of Mendoza, Argentina. The detector consists of about $1660$~surface stations -- water Cherenkov tanks and their associated electronics -- covering an area of $3000~$km$^2$. On dark nights, $27~$telescopes, housed in five fluorescence detector (FD) buildings, are designed to detect air-fluorescence light above the array. The atmospheric properties are continuously monitored during FD data-taking by an extensive and sophisticated atmospheric monitoring system covering the whole observatory~\cite{AugerATMO_LongPaper,MyICRC}. The aerosol component is the most variable term contributing to the atmospheric transmission. It depends on the local environment and long-range transportation. For a deep insight to the aerosol properties at the array of the Auger Observatory, the investigation of air mass origins is needed.

\section{HYSPLIT -- an air-modelling program}
\label{sec:hysplit}
Different air models have been developed to study air mass relationships between two regions. Among them, the {\it HYbrid Single-Particle Lagrangian Integrated Trajectory} model, or HYSPLIT~\cite{HYSPLIT_1,HYSPLIT_2}, is a commonly used air-modelling program in atmospheric sciences that can calculate air mass displacements from one region to another. The HYSPLIT model, developed by the Air Resources Laboratory, NOAA (National Oceanic and Atmospheric Administration)~\cite{websiteNOAA}, computes simple trajectories to complex dispersion and deposition simulations using either puff or particle approaches with a Lagrangian framework. In this work, HYSPLIT is used to get backward/forward trajectories.

Along the trajectory, HYSPLIT requires the wind vector, ambient temperature, surface pressure and height data. These data are available from different meteorological models. The Global Data Assimilation System (GDAS) model provides meteorological data for the site of the Pierre Auger Observatory. They are distributed over a one degree latitude/longitude grid ($360^\circ \times 180^\circ$), with a temporal resolution of three hours. GDAS provides 23 pressure levels, {\it i.e.}\ 23 different altitudes, from $1000~$hPa (roughly at the sea level) to $20~$hPa (height around 26~km). The GDAS grid point closest to the location of the Pierre Auger Observatory is ($35^\circ$ {\bf{S}}, $69^\circ$ {\bf{W}}), {\it i.e.}\ just slightly outside the array to the north-east. Lateral homogeneity of the atmospheric variables across the Auger array is assumed~\cite{AugerATMO_LongPaper}. Validity of GDAS data was previously studied by the Auger Collaboration: the agreement with ground-based weather stations and meteorological radiosonde launches has been verified~\cite{GDASpaper}. The agreement between GDAS model and local measured data has been checked only for state variables of the atmosphere. In the following, the accuracy of the wind data in GDAS are studied, a key parameter in the HYSPLIT model.

\begin{figure*}[!t]
\centerline{{\includegraphics [scale=0.4] {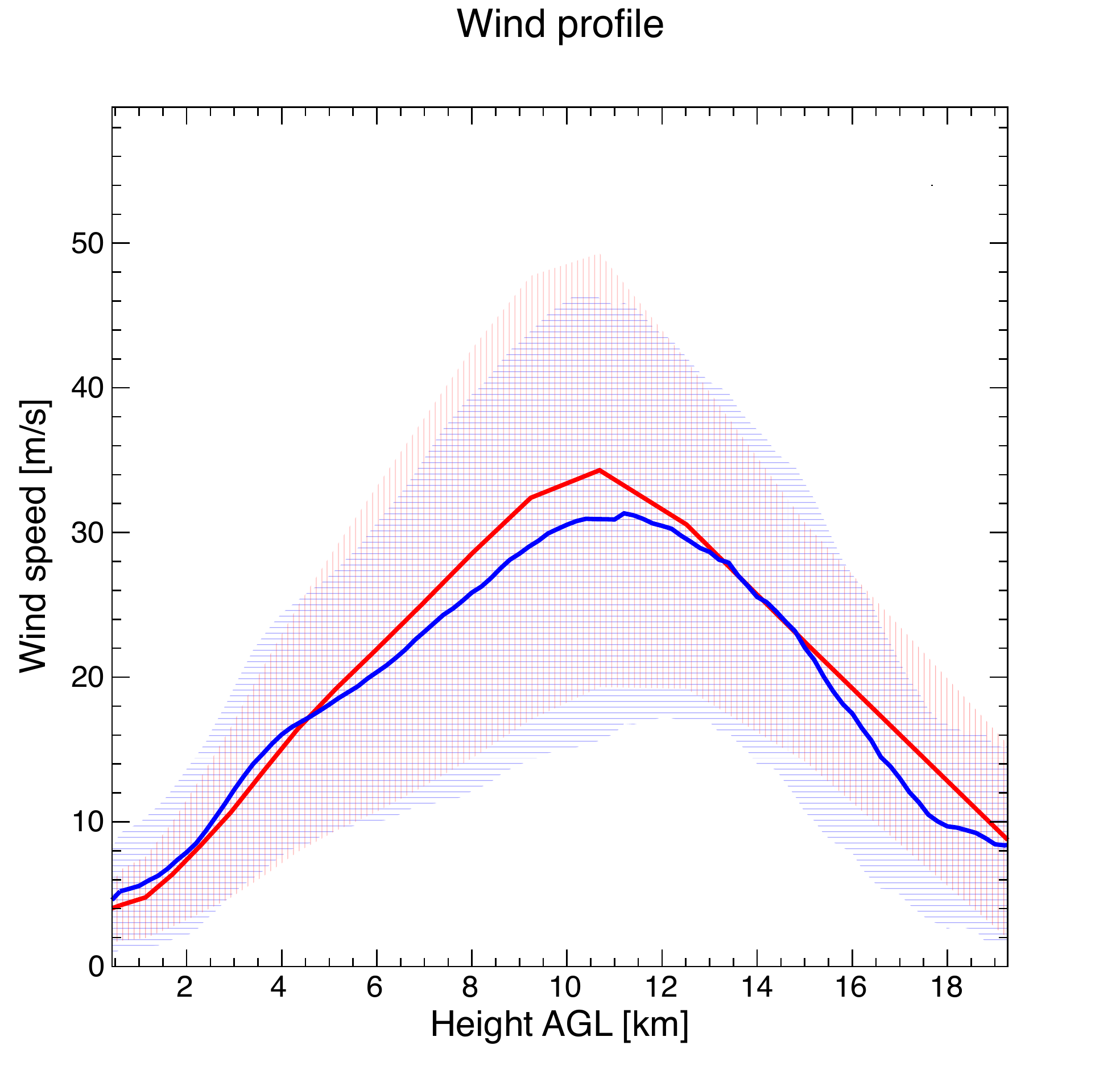}
}
\hfil
{\includegraphics [scale=0.4] {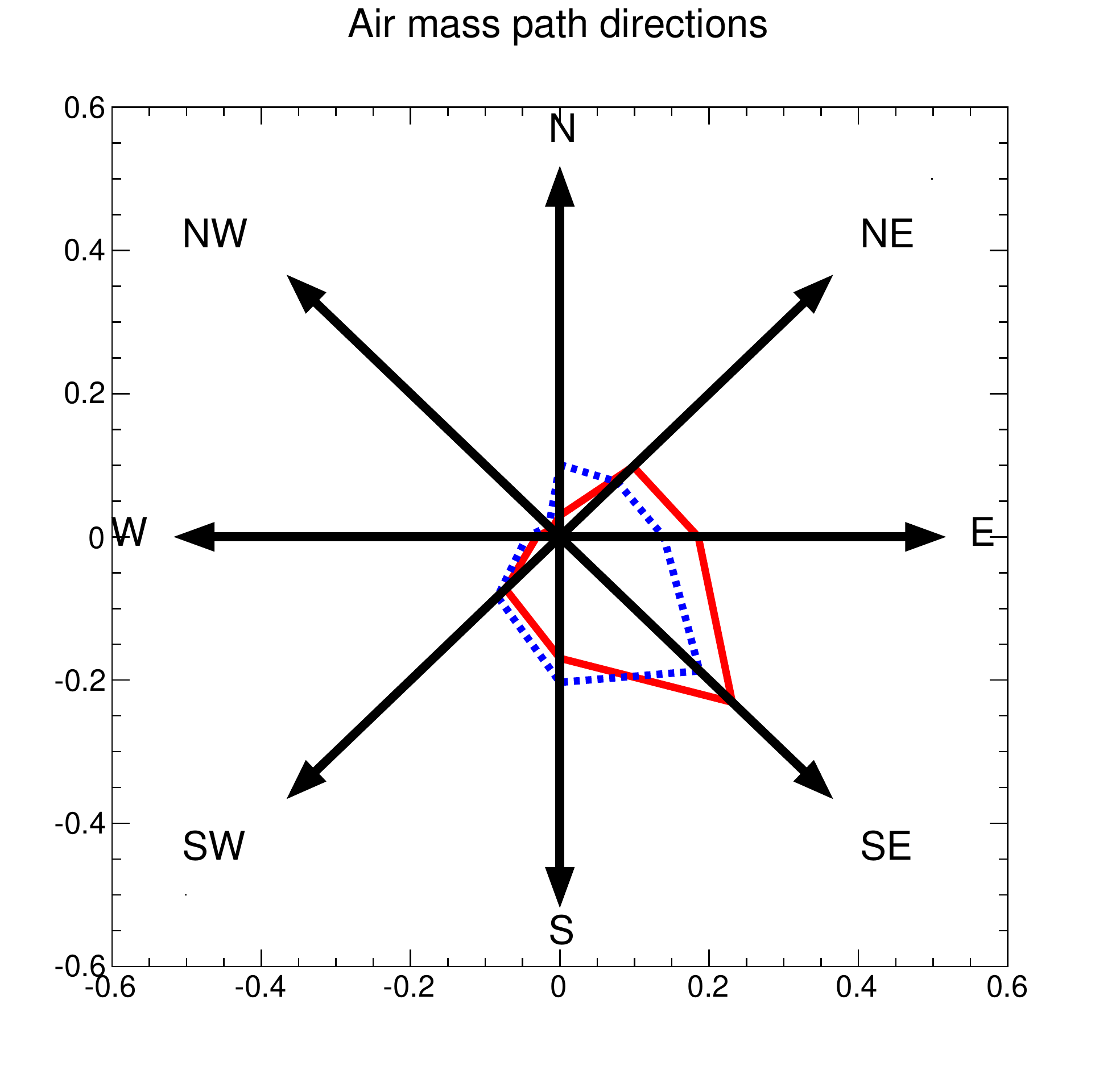}
}
}
\caption{{\bf Comparison between radiosonde data and the GDAS model at the Malarg\"ue location.} Radio soundings from August 2002 to December 2010 and GDAS data from January 2005 to December 2010 are used. {\it Left:} continuous lines represent the averaged wind profile for balloon data ({\it in blue}) and GDAS data ({\it in red}). Bands show the distribution of 68\% of the corresponding samples. {\it Right:} directions of balloon flights are given in {\it dashed line in blue}. The distribution of the air masses traveling above the observatory using HYSPLIT is plotted in {\it solid line in red}.}
\label{fig:GDASwind_validity}
\end{figure*}

\section{Validity of the HYSPLIT calculations using meteorological radio soundings}
\label{sec:validity_balloon}
Above the Pierre Auger Observatory, the height-dependent profiles have been measured using meteorological radiosondes. The balloon flight program ended in December 2010 after having been operated 331 times since August 2002~\cite{BiancaBts}. The radiosonde records data every $20~$m, approximately, up to an average altitude of $25~$km above sea level, well above the fiducial volume of the fluorescence detector. In Figure~\ref{fig:GDASwind_validity}~(left), the averaged vertical profiles of wind speed measured by balloon flights at the observatory or extracted from the GDAS model are displayed. The wind speed, fluctuating strongly from day-to-day, reveals for both data sets a maximum of about $30~{\rm m/s}$ at around $10~$km AGL.

In order to validate the wind directions used in HYSPLIT calculations, it is possible to check the agreement between the directions of the balloon flights and the directions of air mass paths estimated using HYSPLIT. In Figure~\ref{fig:GDASwind_validity}~(right), directions of balloon trajectories operated at the Auger site are given, tending roughly South-Easterly. Only the first $20~$min of the flight are used to estimate the direction of a radio sounding. On the other hand, using the HYSPLIT tool, 48-h forward trajectories from an altitude of $500~$m are computed every hour for the year 2008. The resulting distribution of air mass directions is plotted in red. The agreement between balloon trajectories and forward trajectories computed by HYSPLIT is once again very good. After crosschecks of the vertical profiles for wind speed and the directions of air masses traveling above the Auger array, it can be concluded that air mass calculations for the grid point ($35^\circ$ {\bf{S}}, $69^\circ$ {\bf{W}}) are correct.

\section{Interpretation of aerosol measurements using backward trajectory of air masses}
\label{sec:aerosol_vaod}

\begin{figure}[!t]
\centerline{{\includegraphics [scale=0.42] {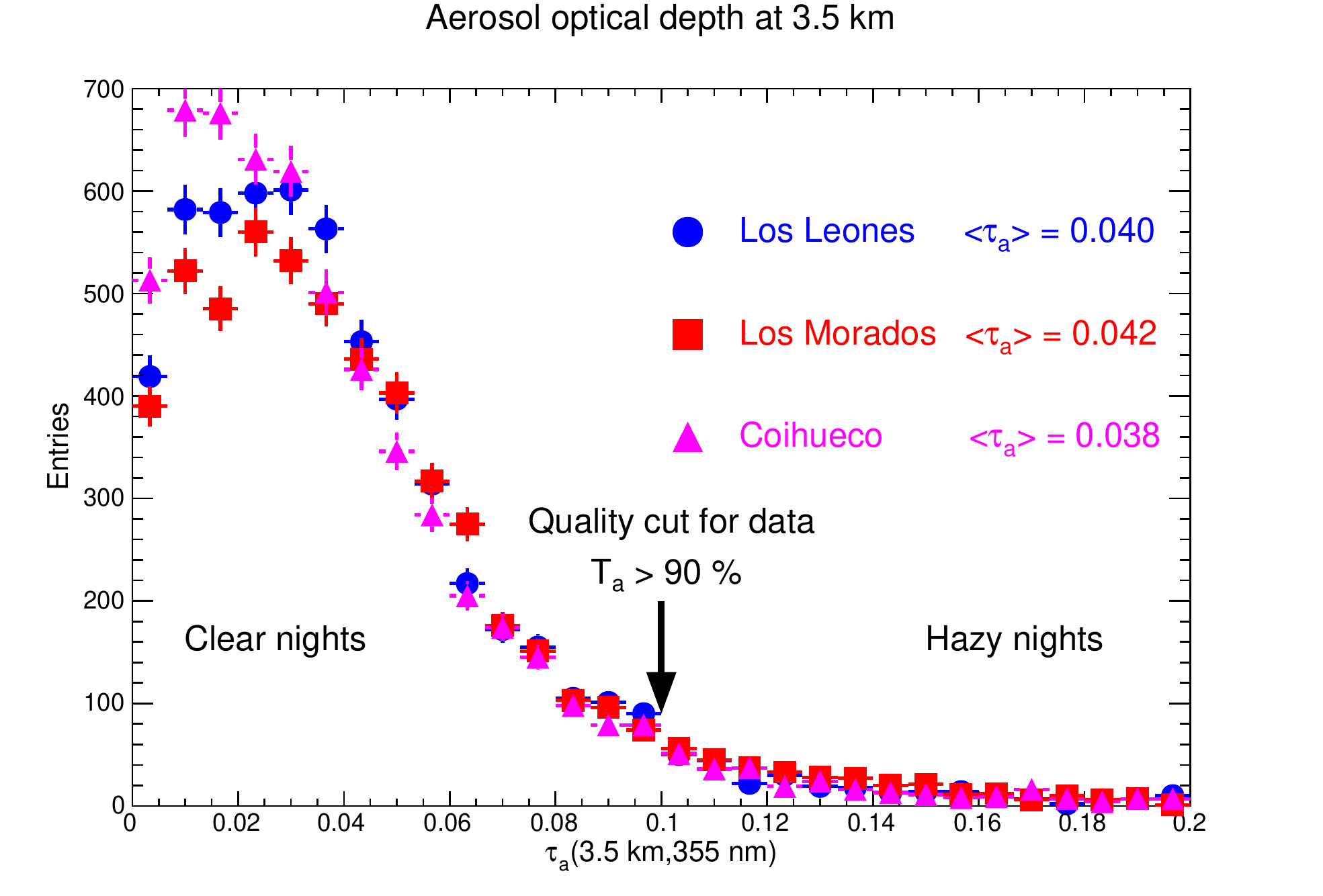}
}
\hfill
{\includegraphics [scale=0.42] {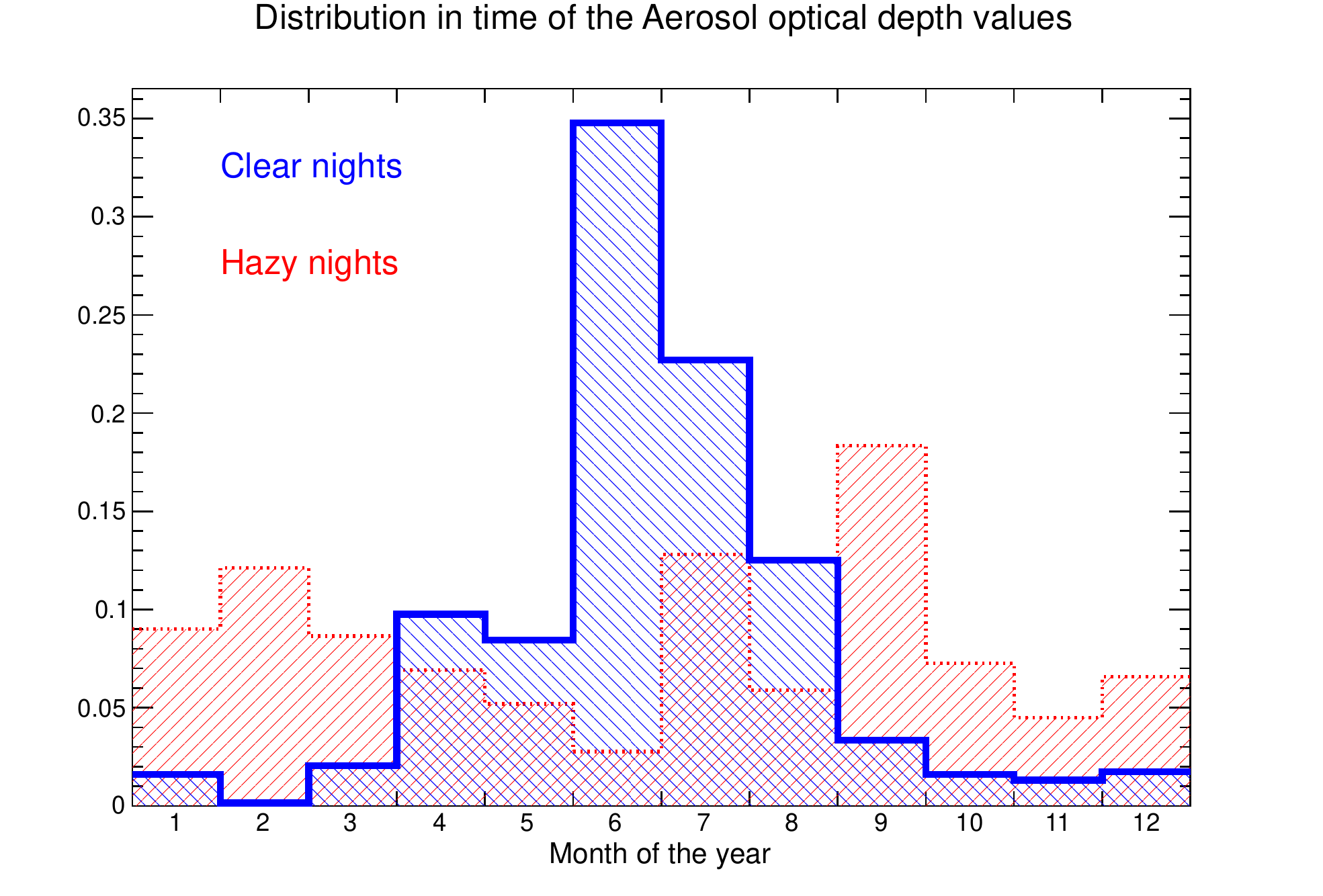}
}
}
\caption{{\bf Aerosol optical depth measurements from the CLF. } {\it Left:} aerosol optical depth at $3.5~$km above the fluorescence telescope stations at Los Leones, Los Morados and Coihueco. Data set between January 2004 and December 2010. {\it Right:} monthly frequency of clear nights ($0.00\leq \tau_a\leq0.01$, {\it continuous line}) and hazy nights ($\tau_a\geq0.10$, {\it dotted line}) at Los Morados.}
\label{fig:VAOD_values}
\end{figure}

At the Pierre Auger Observatory, several facilities have been installed to monitor the aerosol component in the atmosphere. One of the locally aerosol measurements is that of the aerosol optical depth using the Central Laser Facility (CLF)~\cite{CLF_jinst}. It is located at the approximate centre of the Auger surface detector. The main component is a laser producing a vertical beam with a wavelength $\lambda_0$ fixed at $355~$nm, {\it i.e.}\ in the middle of the nitrogen fluorescence spectrum emitted by air showers~\cite{ReviewFLUO}. When a laser shot is fired, the fluorescence telescopes collect a small fraction of the light scattered out of the laser beam, dependent on atmospheric properties. Thus, the aerosol optical depth $\tau_{\rm a}(h,\lambda_0)$ can be estimated, where $h$ is the altitude above ground level~\cite{CLF_valore}. The transmission coefficient is defined as $T_{\rm a} = \exp \,(-\tau_{\rm a})$. Figure~\ref{fig:VAOD_values}~(left) shows the distribution of the aerosol optical depth at $3.5~$km above ground level, recorded at Los Leones, Los Morados and Coihueco fluorescence sites. The mean value for $\tau_a(3.5~{\rm km})$ is around 0.04.

\begin{figure*}[!t]
\centerline{{\includegraphics [scale=0.37] {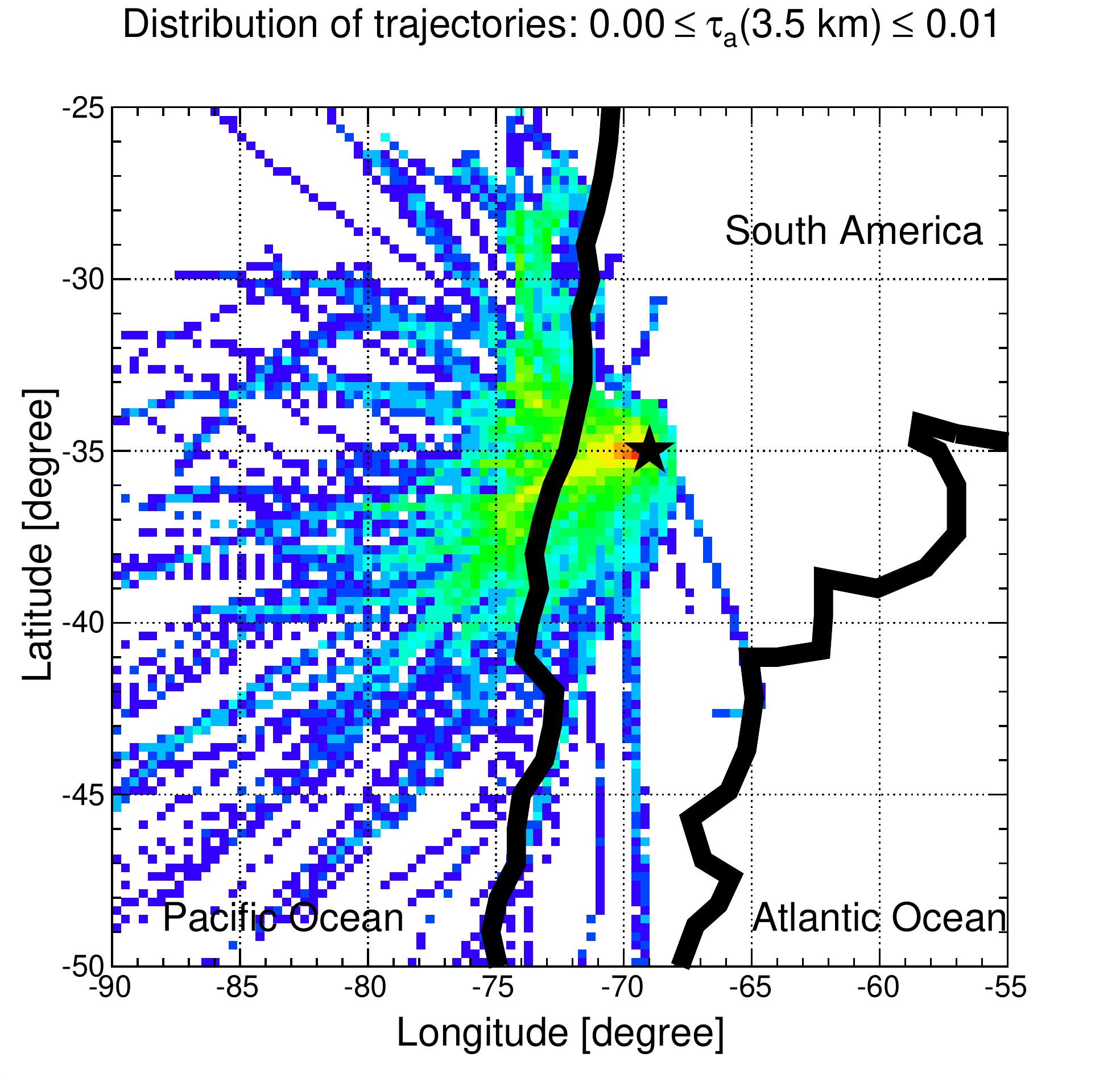}
}
\hfil
{\includegraphics [scale=0.37] {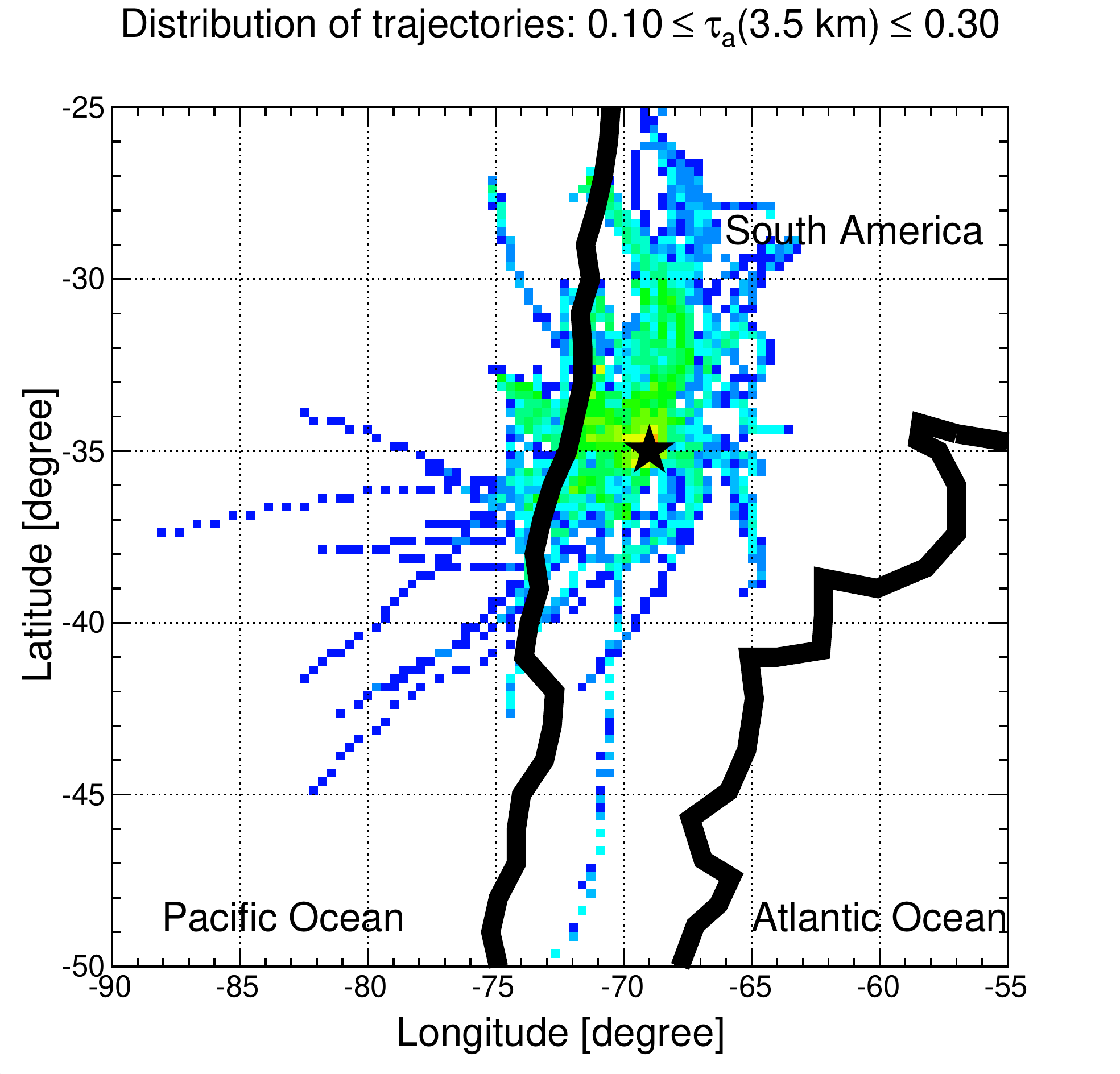}
}
}
\caption{{\bf Distribution of backward trajectories of air masses for clear nights (left) and hazy nights (right).} Paths estimated with HYSPLIT from 2007 to 2010 and aerosol optical depth data coming from the CLF measurements. The black star and the black line represent the Pierre Auger Observatory and the coast, respectively.
}
\label{fig:HYSPLIT_trajectories_VAOD}
\end{figure*}

The data sample of aerosol optical depth measurements at Los Morados site is divided into three sub-sets: {\it clear nights} with the lowest aerosol concentrations ($\tau_a\leq0.01$), {\it hazy nights} with the highest aerosol concentrations ($\tau_a\geq0.10$), and average nights. The relative monthly number of clear and hazy nights is shown in Figure~\ref{fig:VAOD_values}~(right). Clear nights are more common during the austral winter than the rest of the year. The trajectories of air masses arriving at the Auger Observatory have been evaluated for 4 years (2007--2010) using HYSPLIT. A backward trajectory is computed over $48~$hours every hour, throughout the year. The start height is fixed at 500~m above the Malarg\"ue location. The distributions of the backward paths for clear nights and hazy nights are shown in Figure~\ref{fig:HYSPLIT_trajectories_VAOD}: air masses originate mainly from the Pacific Ocean during the clear nights and travel principally through continental areas during the previous 48~hours for hazy nights. Following the conclusion of a chemical aerosol analysis done at the Auger site~\cite{MariaI_LongPaper}, the aerosols originate mainly from the soil. Thus 48-h backward trajectories traveling mainly over the ocean can be characterised as air masses with a low concentration of soil aerosols.

\section{Conclusion}
Using the HYSPLIT tool for tracking air masses from one region to another, a better understanding of air mass behaviour affecting the observations of the Pierre Auger Observatory has been presented: air masses do not have the same origin throughout the year. This difference in origin affects the aerosol concentration: clear nights have air masses coming much more directly from the Pacific Ocean.




\end{document}